# Bio-Based Polyether from Limonene Oxide Catalytic ROP as Green Polymeric Plasticizer for PLA

*Valentina Sessini, Miguel Palenzuela, Jesús Damián, and Marta E. G. Mosquera\**

Departamento de Química Orgánica y Química Inorgánica, Instituto de Investigación Química "Andrés M. del Río". Campus Universitario, E-28871 Alcalá de Henares, Spain.

Corresponding Author e-mail: martaeg.mosquera@uah.es

**ABSTRACT**

In this work, the polymerization of Limonene Oxide (LO) has been achieved using an Earth abundant metal-based catalyst developed in our group, that is very active in ring opening polymerization (ROP) processes. The bio-based polylimonene ether (PLO) obtained had low molecular weight and good thermal properties, thus being a potential green polymeric additive for other bio-based polymers such as PLA. Hence, we have explored its ability to influence PLA properties. The addition of only 10 wt %, led to the modification and improvement of PLA properties in terms of flexibility, thermal stability, and hydrophobicity. The results obtained are promising and open up the potential industrial application of polylimonene oxide (PLO) for the melt-processing of blends based on PLA/PLO. These new materials are totally based on renewable sources and may be interesting for many applications where biodegradability and reduced water adsorption is required, such as food packaging or agricultural mulch films.

**INTRODUCTION**

Renewable resources are increasingly used in the production of chemicals, additives, and plastics. The synthesis of green polymers from biomass-derived monomers such as lactide, terpenoids, vegetable oils, epoxidized natural rubber and so on, is gaining more and more attention in recent years, due to the increased demand of sustainable alternatives for petroleum-based polymers [1-2].

Among those renewable feedstocks, terpene derived epoxides are potential monomers for the synthesis of green polyethers and polyesters by catalytic ring opening polymerization (ROP) [3-4]. In particular, limonene oxide (LO) can be easily produced from limonene, an abundant and readily available monoterpene which is extracted from the peel of citrus fruits and is already widely used as a solvent, fragrance, and insecticide [5-6]. However, the ROP of LO requires the use of efficient catalysts able to overcome the kinetic barrier for LO activation that is higher than for terminal epoxides since LO is characterized by an internal, trisubstituted epoxide motif [7].

There are quite a few studies reported on the copolymerization of LO with anhydrides [8-11] and $CO_2$ [7, 12-15], to generate green polyesters and polycarbonates, respectively. Surprisingly, the studies regarding the homopolymerization of LO are very scarce in literature. In fact, there is only one report, in 2012, by Park *et al*. [6] that described LO photoinitiated cationic ring-opening polymerizations using both diaryliodonium salt and triarylsufonium salt photoinitiators. They found that LO display very high reactivity in photoinitiated cationic ring-opening polymerizations, but the monomer also undergoes several side reactions which results in a mixture of low molecular weight polymeric and nonpolymeric products. They reported that these side reactions appreciably limited the use of the monomer for the synthesis of homopolymers. Hence, the formation of polymers with high molecular weights and good mechanical properties from LO is a challenging goal.

However, low molecular polymers can be used as additives. The demand for non-toxic, environmentally friendly, and biodegradable green-plasticizers, for food and medical packaging has been constantly increasing in the past few years. In fact, plasticizers are among the most important functional additives required for the processing of biopolymers, such as polylactic acid (PLA) and starch [16-17].

In particular, PLA is a biodegradable polymer derived from renewable resources with very attractive properties, i.e. excellent mechanical properties, biocompatibility and good transparency but its brittleness and rigidity impede its application in many fields, thus, additives are needed [18-19].

Plasticizer are typically low molecular weight polymers or oligomers which are used to improve flexibility, plasticity and processability of a polymeric matrix [20]. These additives are able to weaken the intermolecular bond forces of the matrix leading to a decrease of its crystallinity degree, glass transition temperature, elastic modulus and brittleness as well as to confer new properties and multifunctionality to the matrix such as hydrophobicity, amongst others [21]. A large number of low molecular weight polymers have been used as potential plasticizers for PLA such as polyadipates [22], epoxidized oils [23-24] citrate esters [25-26], poly(ethylene glycol) [27-28], lactic acid oligomers [29-30] and so on. However, many of these additives are synthetized from petroleum resources contributing to environmental problems. Therefore, to develop new green plasticizer from renewable sources is still an open challenge.

In our previous work [31] we have synthetized new aryloxide aluminium derivatives with very bulky phenoxide ligands [AlXMeOR], OR = 2,6-bis(diphenylmethyl)-4-tert-butylphenoxide, (X = Cl, Me). These compounds are very active catalysts in ROP for epoxides, in fact [AlClMeOR] is the most active catalyst reported for glycidyl methacrylate (GMA). This compound can polymerize GMA by the oxirane group reaching full conversions within minutes to give polymers with high molecular weights.

In this article, we have explored the activity of this highly active aryloxide aluminium derivative, in the catalytic ROP of LO to obtain green and bio-based additives for PLA. These studies have allowed us to synthetize polylimonene oxide (PLO) by catalytic ROP. The capability of the PLO prepared as potential plasticizer and additive for PLA has been evaluated. The miscibility, morphology, thermal properties, and wettability of the obtained blends have also been investigated.

**EXPERIMENTAL SECTION**

**Materials.** AlClMe$_2$ (0.9 M in heptane) and (+)-Limonene oxide (43.5 % cis-isomer and 56.5 % trans-isomer, calculated by NMR analysis) were purchased from Sigma Aldrich. Poly(lactic acid) (PLA) 3051D, with a specific gravity of 1.25 g cm$^{-3}$, a molecular weight ($M_n$) of ca. 1.42 x 10$^4$ g mol$^{-1}$, and a melt flow index (MFI) of 7.75 g 10 min$^{-1}$ (210 °C, 2.16 kg) was supplied by Nature Works®, USA.

**Polymerization procedure of (+)-Limonene**. The monomer was purified by vacuum distillation using $CaH_2$ as drying agent. Once purified, it was stored at −20 °C under Ar and in the absence of light.

The catalyst (0.011 g, 0.02 mmol) was dissolved in the (+)-Limonene oxide (0.818 ml, 5.0 mmol) in the glovebox. The polymerization was done in bulk, under inert ambient and magnetic stirring at 130 °C and at different reaction times (3, 6, 9, 12, 15, 22, 30 min, 1, 2, 3.5 and 7 h). After the required time, the reaction was quenched at room temperature. At the end of the polymerization one aliquot was taken and quenched with wet $CDCl_3$ to determine the conversion of *cis*-(+)-Limonene in polymer by $^1$H-NMR. The conversion was determined by the integration of the signal of the monomer *cis* versus monomer *trans*, in comparison to the ratio in the starting monomer $^1$H NMR spectrum.

The residual unreacted monomer and active catalyst could influence the macroscopic properties of the polymer obtained. Thus, the polymeric samples were purified by washing the remaining monomer and catalyst using acidified methanol (1 M solution of HCl in methanol) and the precipitated polymers were filtrated by standard Buckner filtration. Finally, they were dried in a vacuum oven at 40°C overnight. The final product was characterized by NMR and IR spectroscopies (general procedures are reported in Supporting Information).

**PLA/PLO film preparation**. PLA and PLO were blended by dissolving the polymers in chloroform. PLO, used as a green additive, was added to the PLA matrix at three different concentrations, 10, 20 and 30 wt % relative to the mass of PLA. The thin films were obtained by solvent casting method in a glass petri dish and evaporated at room temperature under ventilation for 24 h to obtain films with a thickness of about 100 μm. The obtained samples were named as PLA10PLO, PLA20PLO and PLA30PLO for the three different blends, thus emphasizing the respective PLO content. An unplasticized sample was prepared in the same way for comparison and it was named PLA.

**Thermal and morphological analysis.** The thermal characterization was performed by dynamic differential scanning calorimetry (DSC) analysis and by thermogravimetric analysis (TGA). For DSC measurements a Mettler Toledo DSC822e instrument was used performing heating/cooling/heating cycles program in the range of -40 to 200 °C with a heating/cooling rate of 10 °C/min and run under nitrogen purge (30 mL/min). The glass transition temperature ($T_g$) was calculated from the second heating scan and was taken at the mid-point of heat capacity changes. The melting temperature ($T_m$) and cold crystallization temperature (T$_{cc}$) were obtained from the second heating, and the degree of crystallinity ($\chi_c$) was determined by using Equation (1):

$$\chi_c = 100 \times \left[\frac{\Delta H_m - \Delta H_{cc}}{\Delta H_m^{100}}\right] \frac{1}{1-m_f} \qquad \text{Equation (1)}$$

where ΔH$_m$ is the enthalpy of fusion, ΔH$_{cc}$ is the enthalpy of cool crystallization, $\Delta H_m^{100}$ is the enthalpy of fusion of a 100 % crystalline PLA, taken as 93 J/g. and *1- m$_f$* is the weight fraction of PLA in the sample.

The TGA measurements were carried out using a TA-TGA Q500 analyzer. The experiments were performed using about 10 milligrams of sample from room temperature to 700 °C at 10 °C/min under nitrogen atmosphere (60 mL/min). The morphology of the samples previously sputter-coated with gold was studied by a Scanning Electron Microscopy (ZEISS DSM-950 instrument) operating at 25 kV. The wettability and hydrophobicity of the samples surface was tested using an optical contact angle meter (Krüss DSA25 Drop Shape Analysis System) operating at room temperature. Samples were placed on the test cell and drops of distilled water were placed on the surfaces by the delivering syringe.

## RESULTS AND DISCUSSION

We performed the polymerization reaction of (+)-limonene oxide (LO) using [AlClMeOR] as catalyst, in bulk at 250:1 Mon:Cat ratio and checked the evolution at different times (Scheme 1). The aluminium compound proved to be a very active catalyst for the ROP and was able to polymerize LO in 30 minutes.

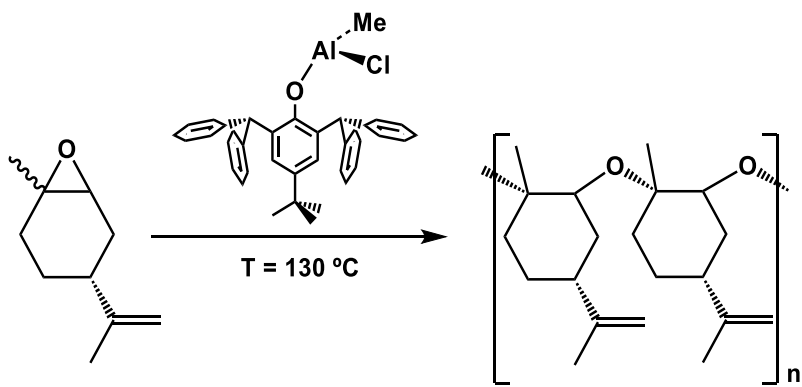

Scheme 1. Catalytic polymerization of (+)-limonene oxide.

In the reaction only the *cis* monomer was polymerized, while the *trans* isomer was left intact. This behaviour is somehow opposite to the one described by Coates and others in the copolymerization processes where only the *trans* isomer reacts [9, 15].

The formation of PLO was confirmed by IR spectroscopy and $^1$H-NMR. In Figure 1, the $^1$H-NMR spectra for the monomer and the polymer obtained after 30 min of reaction are compared. As shown in the figure, the signals due to the protons from the epoxide group have disappeared while a signal centered at 3.5 ppm corresponding to the ether linkages of the polymer is observed. The doublet at 2.98 ppm for the LO *trans*-isomer remains unchanged (Figure 1, middle spectrum). The unreacted *trans* isomer was eliminated by washing the reaction mixture with acidic MeOH (Figure 1, bottom spectrum).

In the FTIR spectra (Figure S1) the band for the epoxide ring vibration [32] of the LO monomer, that should appear at 842 cm$^{-1}$, is missing in the PLO spectrum while one new intense band associated with the ether linkage at 1068 cm$^{-1}$ appears, suggesting that the LO polymerization has occurred through the opening of the epoxide ring.

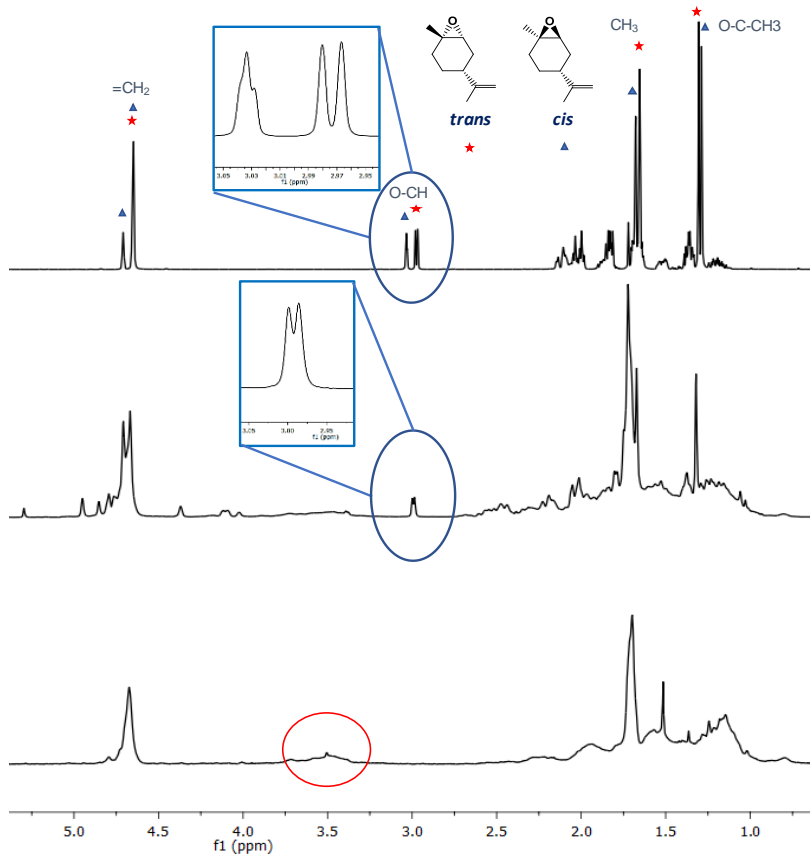

Figure 1. Comparison of NMR spectra from the top to the bottom: (+)-LO monomer, PLO before washing and PLO washed.

Similar results were obtained from other authors for the homopolymerization of 1, 2-cyclohexene oxide [33-34]. Moreover, in the region comprised in the 3600-3200 cm$^{-1}$ range it is possible to identify the band related with the OH terminal groups of PLO.

The polymers obtained have low molecular weights, up to 1300 Da, and moderate polydispersities (1.37-1.42) (table S1). We checked the kinetics of the polymerization by analyzing the products mixtures at given reaction times. The plot of the ln($M_0/M_t$) *vs* time is linear which agrees with a polymerization process with first order dependence on the monomer concentration (Figure 2).

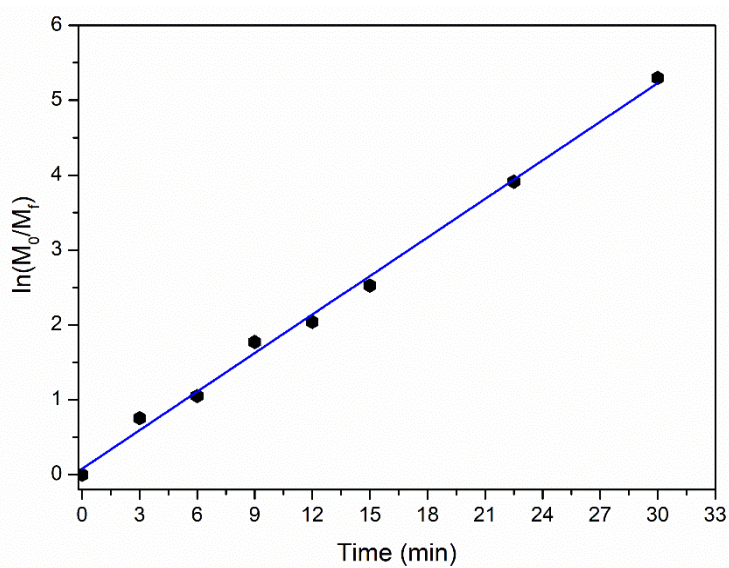

Figure 2. Kinetic plot for ROP of LO by [AlClMeOR].

However, once full conversion of the *cis* isomer is reached, after 30 minutes, if the reaction was left for longer no increase of the polymer molecular weight was observed. The limitation in the $M_n$ obtained could be attributed to the presence of charge transfer agents (CTA) generated in lateral reactions. Several side reactions have been reported for LO in the presence of Lewis acids, such as the rearrangement described by Duchateau to a ketone and an allylic alcohol shown in Figure 3 [10, 35]. Furthermore, LO can also react with traces of water present in the reaction media that may lead to the formation of the *trans* diaxial diol [36].

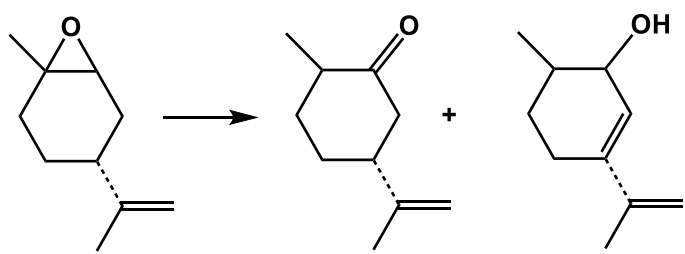

Figure 3. Rearrangement reaction of limonene oxide.

A MALDI-TOF study showed two major types of endcaps for the polymers (Figure S3), the most abundant one would agree with the allylic alcohol derived from the rearrangement of the LO (Figure 3). As such, a signal at m/z= 783.4 Da is found and matches four units of the monomer and the capping allylic alkoxyde group, (Figure S4).

The other end group includes a water molecule which could account for the diol coming from the reaction of the LO with water, peak at m/Z = 801.4 Da, (Figure S5).

In the ROP of other epoxides using [AlClMeOR] as catalyst the mechanism observed was coordination insertion, and when no other co-initiator is used, as in this case, the end group was the phenol ligand, as detected by mass spectra [31]. For the LO polymerization described herein, the phenol ligand was not detected as end group so a mechanism where the initiation occurs through coordination of the epoxide to the aluminum due its Lewis acid character, followed by the addition of monomer molecules to the complex, as proposed by R. Bacskai [33], maybe taking place. The nature of the termination step is not clear but since different end-groups are observed it can be suggested that there is more than one path for the polymerization ending.

Aiming to apply the synthesized PLO as a new bio-based and environmental-friendly additive for blending with PLA and improve its properties, we chose the PLO obtained after 30 min of reaction since it showed complete conversion of LO *cis*-isomer and it has the lower polydispersity (1.37).

The miscibility of a blend and consequently the phase behavior are key parameters to define its compatibility and how the properties of the blended polymers can change. Thus, when an additive is applied to a polymeric matrix, the study of the miscibility of this new blend is required as a crucial characterization to forecast the possible improvement of the matrix properties. There are many techniques to characterize the miscibility of a polymeric blend. Differential scanning calorimetry (DSC) is one of the most widely used since it is able to indicate the number of phases present in a blend detecting the $T_g$ of the studied system [37]. The miscibility level can be deduced by means of the shift of $T_g$ of each phase compared to those of the neat component [38].

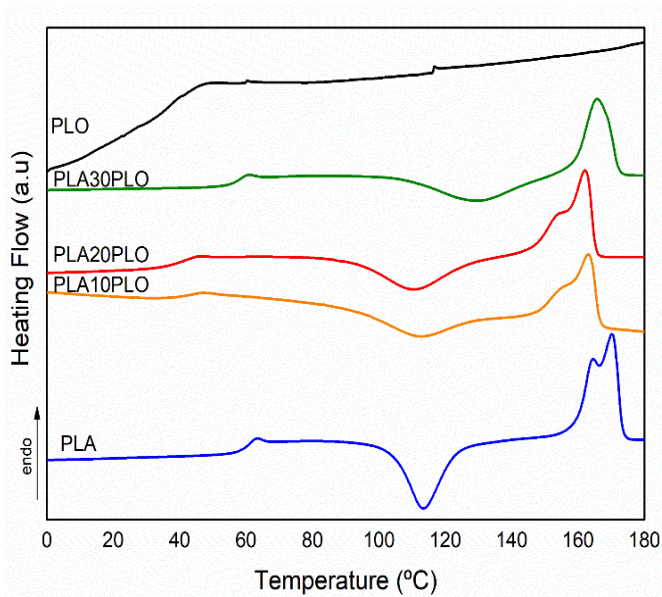

Figure 4. DSC thermogram reporting the second heating scan of neat PLA and PLO and their different blends.

Figure 4 displays the DSC thermograms of the second heating scan for the neat polymers as well as the different blends. As expected, the PLO curve showed an amorphous behaviour typical of low molecular weight polymers, with a $T_g$ at 37 °C. The PLA curve and those of its blends showed similar behaviour and shows the typical thermal properties of PLA. The $T_g$ of neat PLA is 60 °C while the exothermic peak in the 90 °C-130 °C range corresponds to the cold crystallization process with a maximum crystallization rate at 113 °C ($T_{cc}$). Finally, the endothermic peak between 155 °C to 175 °C corresponds to the melting process of the crystalline phase in PLA showing a maximum at 170 °C ($T_m$).

The thermal properties of all samples are reported in Table 1. The DSC curves showed for all the blends a single $T_g$ which is in agreement of a good miscibility between PLO and PLA. Moreover, a significant reduction in $T_g$ values of PLA was observed due to the increase in the polymer chain mobility, induced by the presence of PLO that acts as plasticizer. A decrease of about 15 °C was registered for PLA10PLO with respect to the neat PLA, while for PLA20PLO the value remained constant. This behaviour was probably caused by a low interaction between the two polymers in the blend and the phase separation of PLO at concentrations higher than 10 wt %, reaching the limit of the miscibility.

Table 1. Thermal properties and WCA for all the samples.

| Sample | $T_g$(°C) | $T_{cc}$(°C) | $T_m$(°C) | $T_{5\%}$(°C) | $T_{max}$(°C) | WCA (°) |
|---|---|---|---|---|---|---|
| **PLO** | 37 | - | - | 166 | 189, 385 | - |
| **PLA** | 60 | 113 | 170 | 308 | 357 | 77 ± 4 |
| **PLA10PLO** | 41 | 113 | 163 | 297 | 360 | 88 ± 2 |
| **PLA20PLO** | 42 | 110 | 162 | 289 | 350 | 86 ± 1 |
| **PLA30PLO** | 58 | 129 | 166 | 289 | 342 | 87 ± 3 |

Similar behaviour has been observed by different authors using 10 wt % of polyethylene glycol (1,000 g mol$^{-1}$) [39-40], 20 wt % of triblock-copolymers based on pectinate polymethylphenylsilane-co-poly(ε-caprolactone) [41] and 20 wt % of isosorbide dioctoate [42]. Furthermore, when 30 wt % of PLO were added to PLA, the $T_g$ increased reaching a value similar to that of the neat matrix, confirming the phase separations of PLO. The same behaviour was observed for the $T_{cc}$ and $T_m$ values, a decrease is observed for the samples with 10 and 20 wt % of PLO and the maximum saturation is reached for PLA30PLO. The presence of PLO acting as plasticizer for PLA provokes an increase of mobility of the PLA chains favoring the crystallization process at lower temperatures as it has been reported previously for PLA-limonene systems [43]. However, the calculation of the crystallinity degree shows that all the samples are totally amorphous ($\chi_c$<5%).

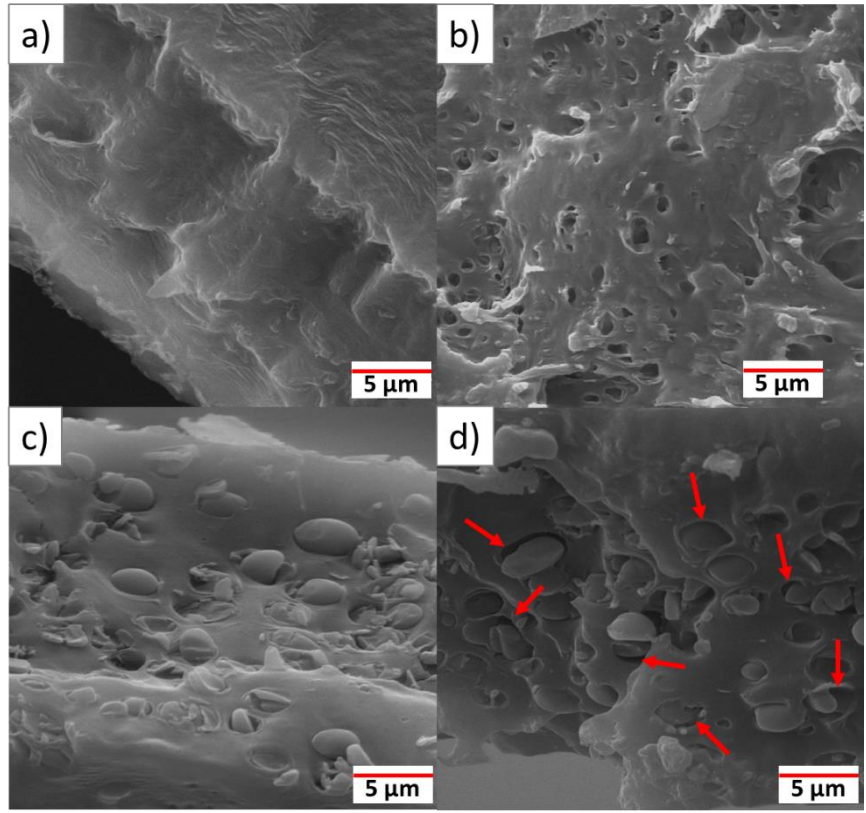

Figure 5. SEM micrographs of a) neat PLA, b) PLA10PLO, c) PLA20PLO and d) PLA30PLO.

The morphologies of the cryo-fractured surface of the blends were also analyzed in order to evaluate the thermodynamic miscibility and the phase separation by SEM analysis. As shown in Figure 5, neat PLA displayed a rough surface revealing its typical brittle fracture. On the other hand, when 10 wt % of PLO were added, few numbers of empty micro voids were observed indicating the coalescence of the PLO-rich phase into microspheres dispersed into the PLA matrix (Figure 5b) revealing a partial phase separation. When the amount of PLO was increased in the blend, the dimension of PLO-rich phase spheres increased as well, indicating that the interactions between PLO/PLO-rich phases is higher than those between PLO/PLA phases, thus leading to a more pronounced phase separation (Figure 5c). Confirming this behaviour, in the PLA30PLA micrograph (Figure 5d), it is possible to observe a void space between PLA and PLO dispersed phases indicating an evident decreasing of the miscibility of the blend and the absence of strong interaction between the two phases in contact.

These results confirm the thermodynamic miscibility of the blend PLA/PLO, previously discussed, although they highlight a partial phase separated morphology composed by a PLO-rich phase dispersed into the PLA matrix that was not detectable by DSC measurements. Similar results in terms of thermal properties and morphology were previously observed for other systems in literature [24, 41-42]. Additionally, the visual appearance of the blends (Figure S7) indicates the phase separation also at a macroscopic level. Indeed, the addition of PLO into the matrix reduces PLA transparency.

Thermogravimetric analysis (TGA) experiments were also performed to investigate the thermal stability of the neat polymers and their blends, the results are showed in Figure 6. In Table 1, the results in terms of initial temperature of degradation at 5 % of weight loss as well as the maximum temperature of degradation are reported. PLO decompose in a two-step process with an initial degradation of 166 °C and the maximum degradation rate temperatures centered at 189 and 385 °C, respectively. The addition of PLO to PLA, slightly decreased the thermal stability of the polymeric matrix. The initial decomposition temperature of PLA with the different concentrations of PLO is lower compared to that of neat PLA, reaching the lower value for PLA20PLO and PLA30PLO at around 290 °C. Moreover, the addition of 10 wt % of PLO to PLA matrix did not affect the $T_{max}$, which was maintained around 360 °C. However, for a higher amount of PLO the $T_{max}$ decreased of about 10 and 20 °C for PLA20PLO and PLA30PLO, respectively. Similar results were obtained for PLA/limonene systems [43]. These results show that PLO is stable within the range of PLA melting processing, opening its application as an industrial plasticizer.

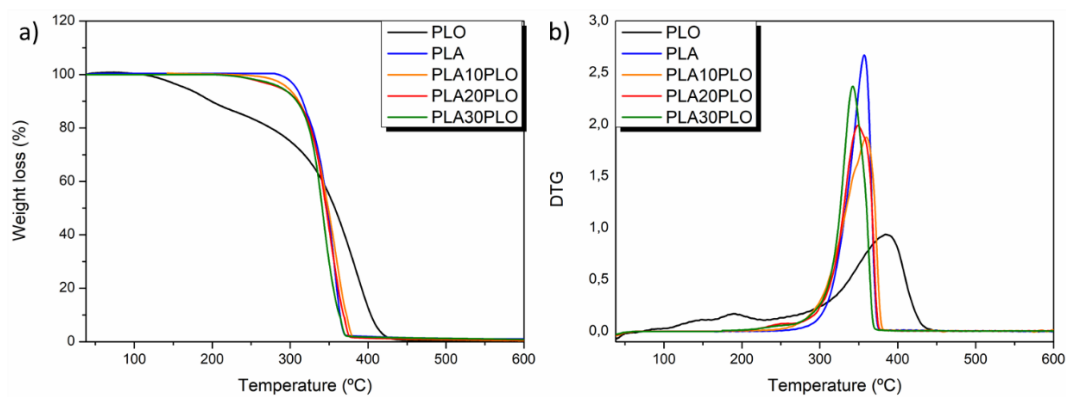

Figure 6. TGA (a) and DTG (b) images for the neat polymers and their different blends.

The water contact angle (WCA) of neat PLA film and its blends with PLO was determined in order to evaluate their wetting behavior and to verify if PLO could modify PLA properties. As observed in Table 1, the contact angle in all films is greater than 65°, which represents a surface with a hydrophobic behavior according to Vogler *et al.* [44]. However, it is observed that the incorporation of 10 wt % of PLO in the PLA film provides a slight increase in its hydrophobicity from 77° to 88° due to the hydrophobic character of PLO (Figure 7).

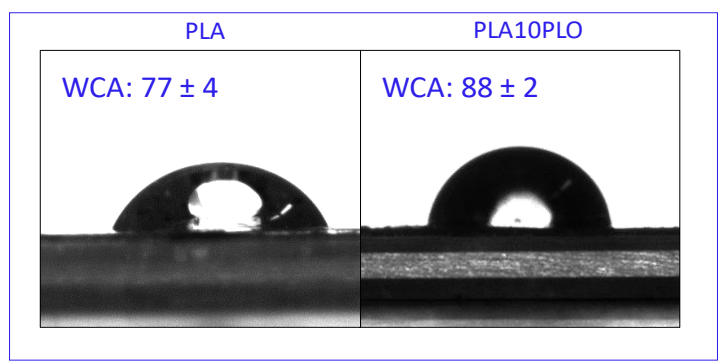

Figure 7. Pictures of PLA and PLA10PLO films with the water drop used to determine the contact angle.

Similar results were found previously by Arrieta *et al.* [45] for PLA/Limonene blends, reporting an WCA increase of about 20°. The increase of WCA is prominent up to 10 wt % of PLO in the blend, for higher amounts, the WCA value stays constant. Due to their increased hydrophobicity, PLA/PLO blends are also expected to have higher resistance to water adsorption than neat PLA. Thus, these new materials may be interesting for many applications where reduced water adsorption is required.

FTIR spectra were recorded for neat PLA and PLO as well as for their blends in order to investigate the chemical structure of the films and to study their characteristic peaks (Figure 8).

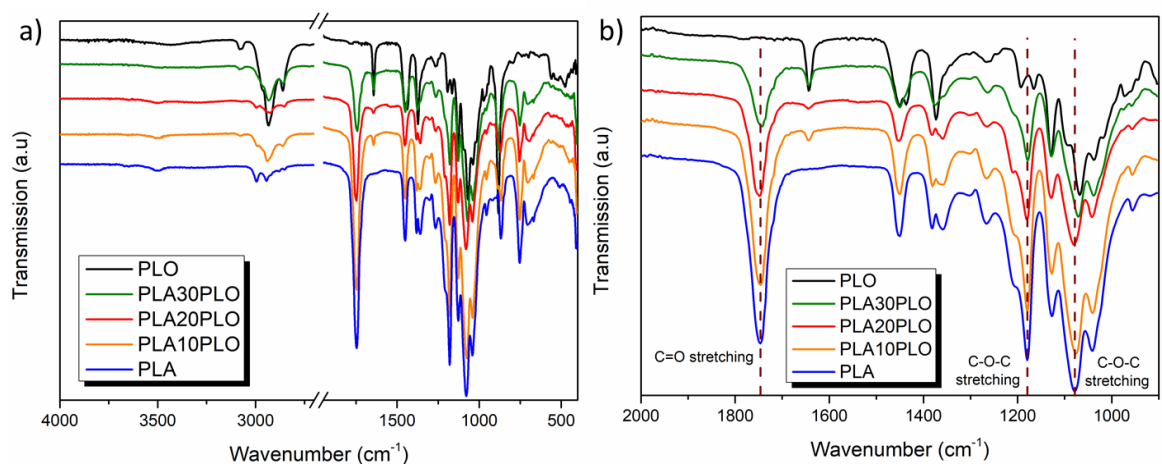

Figure 8. FTIR spectra of neat materials and their different blends: (a) in the 4000 -400 cm$^{-1}$ and (b) 2000-900 cm$^{-1}$ regions.

PLA shows the stretching frequencies for C=O, asymmetric and symmetric –CH$_3$, at 1746, 2997 and 2943 cm$^{-1}$ respectively. At 1180 cm$^{-1}$ it is possible to observe the band related with the O-C-O groups as well as three peaks at 1128, 1081 and 1042 cm$^{-1}$ belonging to the C-C-O groups. Moreover, bending frequencies for –CH$_3$ have been identified at 1452, 1380 and 1359 cm$^{-1}$ [28, 43].

Meanwhile, PLO shows the characteristic stretching frequency of C=C at 1642 cm$^{-1}$ while the deformation frequency of CH$_2$ and CH$_3$ were detected at 1452, 1436 and 1373 cm$^{-1}$, respectively. Furthermore, the band related to the C-O group was observed at 1067 as well as a broad peak at 3434 cm$^{-1}$, which corresponds to terminal hydroxyl group of PLO [32]. PLA/PLO systems showed almost the same transmission peaks as neat PLA but the peak corresponding with the C-O group of PLO was shifted throughout higher wavenumber, from 1067 to 1077 and 1081 cm$^{-1}$, for PLA10PLO and PLA20PLO, indicating a possible chemical interaction between the polymers. These results are in good agreement with those obtained with the other technics, where the lower amount of PLO, 10 wt %, was interacting with the PLA matrix and was able to modify and improve PLA properties in terms of flexibility, thermal stability and hydrophobicity.

**CONCLUSIONS**

The aluminium ROP catalyst [AlClMeOR], (OR = 2,6-bis(diphenylmethyl)-4-*tert*-butylphenoxide) developed in our group is an active catalyst for the polymerization of LO, which allowed us to obtain poly limonene oxide totally derived from renewable sources. The polyether obtained had a low molecular weight and good thermal properties being a promising green additive for other bioplastics. When testing its ability to modify PLA, just with a blend of 10 wt %, PLO was able to modify and improve PLA properties in terms of flexibility, thermal stability, and hydrophobicity. The results showed that PLO is stable within the range of PLA melting processing and opens its application as industrial plasticizer. Moreover, due to their increased hydrophobicity, PLA/PLO blends are also expected to have higher resistance to water adsorption than neat PLA. Thus, these new materials may be interesting for many applications where reduced water adsorption is required. Further studies on the influence of the microstructure of the PLO are ongoing as well as on the mechanical properties of the blend.

**ASSOCIATED CONTENT**

Additional information regarding experimental details and physical data (General procedures and Synthesis of [AlMeCl(2,6-(CHPh$_2$)$_2$-4-$^t$Bu-C$_6$H$_2$O)]); Polymer characterization: General conditions, percentage of conversion, molecular weight and polydispersity (Table S1), FTIR spectrum of (+)-Limonene Oxide monomer compared with PLO (Figure S1), HSQCed $^1$H-$^{13}$C NMR spectrum of PLO (Figure S2); MALDI-TOF mass spectra of PLO (Figure S3) and polymeric chain for the predominant series at 479 and 497 Da (Figure S4 and Figure S5); Visual appearance of the obtained films (Figure S6).

# AUTHOR INFORMATION

**Corresponding Author**

Marta E. G. Mosquera, martaeg.mosquera@uah.es

**Author Contributions**

The manuscript was written through contributions of all authors. All authors have given approval to the final version of the manuscript.

# ACKNOWLEDGMENT

Financial support from Spanish Government [RTI2018-094840-B-C31] and the Alcalá University, Spain [CCG2018/EXP-038] are gratefully acknowledged. M.P. thanks the Universidad de Alcalá for a Predoctoral Fellowship. Authors thank Dr. L. Peponi and the Institute of Polymers Science and Technology, ICTP-CSIC, for the thermal analysis. This project has received funding from the European Union's Horizon 2020 research and innovation programme under the Marie Skłodowska-Curie grant agreement No 754382, GOT ENERGY TALENT. The content of this article does not reflect the official opinion of the European Union. Responsibility for the information and views expressed herein lies entirely with the authors.